\documentclass[%
 aip,
 amsmath,amssymb,
 reprint,%
]{revtex4-1}

\usepackage{graphicx}
\usepackage{dcolumn}
\usepackage{bm}

\usepackage[utf8]{inputenc}
\usepackage[T1]{fontenc}
\usepackage{mathptmx}
\usepackage{etoolbox}

\makeatletter
\def\@email#1#2{%
 \endgroup
 \patchcmd{\titleblock@produce}
  {\frontmatter@RRAPformat}
  {\frontmatter@RRAPformat{\produce@RRAP{*#1\href{mailto:#2}{#2}}}\frontmatter@RRAPformat}
  {}{}
}%
\makeatother
\begin{document}

\preprint{AIP/123-QED}

\title{A Spin Field Effect Transistor Based on a Strained Two Dimensional Layer of a Weyl Semimetal}
\author{Rahnuma Rahman}
\author{Supriyo Bandyopadhyay}%
 \email{sbandy@vcu.edu.}
 \homepage{http://www.people.vcu.edu/~sbandy
 }
\affiliation{Department of Electrical and Computer Engineering, Virginia Commonwealth University, Richmond, VA 23284, USA
}%


\begin{abstract}
Spin field effect transistors (SpinFET) are an iconic class of spintronic devices that exploit gate tuned spin-orbit interaction in semiconductor channels interposed between ferromagnetic source and drain contacts to elicit transistor functionality. Recently, a new type of SpinFET based on gate tuned {\it strain} in quantum  materials (e.g. topological insulators) has been proposed and may have interesting analog applications, such as in frequency multiplication, by virtue of its unusual oscillatory transfer characteristic. Here, we propose and analyze yet another type of SpinFET  in this class, which may have a different application. It is based on a Weyl semimetal. Because the operating principle is non-classical, the channel conductance shows {\it oscillatory} dependence on the channel length at zero gate voltage. Furthermore, the transconductance can switch sign if the channel length is varied. This latter feature can be exploited to implement a complementary device like CMOS by connecting two such SpinFETs of slightly different channel lengths in series. These unusual properties may have niche applications.
\end{abstract}

\maketitle

Topological insulators and Weyl semimetals are a new class of spintronic materials with many intriguing properties whose device applications, however, have remained relatively unexplored. Recently, we proposed and analyzed a spin-field-effect-transistor (SpinFET) based on the spin-momentum locked surface states of a three dimensional topological insulator (TI) deposited on a piezoelectric thin film \cite{one}. A gate voltage generates strain in the piezoelectric, which is transferred to the TI channel, and that modulates the quantum interference between the spin states in the TI channel to vary the channel conductance. This implements a transistor that has an unusual oscillatory transfer characteristic which can be exploited to realize a single transistor frequency multiplier \cite{one}. Motivated by that idea, we propose here a similar SpinFET that is based on a two-dimensional layer of a Weyl semimetal and may have different applications. 

In the simplest case, the Hamiltonian of a two-dimensional layer of a Weyl semimetal (in the vicinity of the two Weyl nodes) can be written as \cite{two}
\begin{equation}
H = \hbar v \left [\pm [\sigma_x]\left( k_x \pm k_{x0} \right ) + [\sigma_y]\left( k_y \pm k_{y0} \right )\right ],
\end{equation}
where the two signs correspond to the two Weyl nodes of opposite chirality, $v$ is the velocity of carriers, the [$\sigma$]-s are the Pauli spin matrices, and $k_{x0},k_{y0}$ are constant wave vectors. 

\begin{figure}[!t]
\centering
\includegraphics[width = 0.46\textwidth]{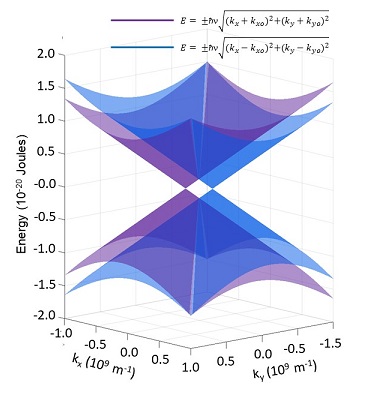}
\caption{Dispersion relations in a two-dimensional Weyl semimetal calculated from Equation (\ref{dispersion}). The material parameters are assumed to be: $v$ = 10$^5$ m/s, $k_{x0} = k_{y0}$ = 10$^8$ m$^{-1}$.}
\label{fig:dispersion}
\end{figure}

Diagonalizing this Hamiltonian, we get the energy dispersion relations as
\begin{eqnarray}
    E & = & \pm \hbar v \sqrt{\left (k_x + k_{x0} \right )^2 + \left (k_y + k_{y0} \right )^2} \nonumber \\
    E & = & \pm \hbar v \sqrt{\left (k_x - k_{x0} \right )^2 + \left (k_y - k_{y0} \right )^2}
    \label{dispersion}
\end{eqnarray}
in the neighborhood of the two nodes. A plot of the dispersion relations ($E$ vs. $kx, ky$) are shown in Fig. \ref{fig:dispersion}. 

In the first valley associated with the first Weyl node in Fig. 1, the two (wavevector-dependent) eigenspinors are given by
\begin{eqnarray}
\psi_{\uparrow}^{(1)} & = &
 \left ( 
 \begin{array}{c}
 1 \\
 e^{i \theta_1}
 \end{array}
 \right ) \nonumber \\
 \psi_{\downarrow}^{(1)} & = &
 \left ( 
 \begin{array}{c}
 1 \\
 e^{-i \theta_1}
 \end{array}
 \right ),
 \label{eigenspinor1}
\end{eqnarray}
where $\theta_1 = arctan \left [ {{k_y + k_{y0}}\over{k_x + k_{x0}}}\right ]$.
In the second valley associated with the second Weyl node, the two eigenspinors are given by 
\begin{eqnarray}
\psi_{\uparrow}^{(2)} & = &
 \left ( 
 \begin{array}{c}
 1 \\
 e^{i \theta_2}
 \end{array}
 \right ) \nonumber \\
 \psi_{\downarrow}^{(2)} & = &
 \left ( 
 \begin{array}{c}
 1 \\
 e^{-i \theta_2}
 \end{array}
 \right ),
 \label{eigenspinor2}
\end{eqnarray}
where $\theta_2 = arctan \left [ {{k_y - k_{y0}}\over{k_x - k_{x0}}}\right ]$.

We will now consider the effect of strain on the Weyl semimetal, which has been examined in, for example,  refs. [\onlinecite{three,four,five,six,seven,eight}]. Some of these effects can only be observed at very high strain which may not be possible to generate electrically. Ref. [\onlinecite{three}] 
modeled the effect of moderate strain as a shift in the wavevectors and energy in the Hamiltonian:
\begin{equation}
H = \delta E_0 + \hbar v \left [\pm [\sigma_x]\left( k_x +\delta k_x \pm k_{x0} \right ) + [\sigma_y]\left( k_y + \delta k_y \pm k_{y0} \right )\right ],
\end{equation}
which will change the energy dispersion relations around the two Weyl nodes to 
\begin{eqnarray}
    E & = & \delta E_0 \pm \hbar v \sqrt{\left (k_x + \delta k_x + k_{x0} \right )^2 + \left (k_y + \delta k_y + k_{y0} \right )^2} \nonumber \\
    E & = & \delta E_0 \pm \hbar v \sqrt{\left (k_x + \delta k_x - k_{x0} \right )^2 + \left (k_y + \delta k_y - k_{y0} \right )^2}.
    \label{new-dispersion}
\end{eqnarray}
The quantities $\delta E_0$, $\delta k_x$ and $\delta k_y$ depend only on the strain $\epsilon$.

The eigenspinors will still be given by Equations (\ref{eigenspinor1}) and (\ref{eigenspinor2}), except now the thetas will be given by 
$\theta_1 = arctan \left [ {{k_y + \delta k_y (\epsilon)  + k_{y0}}\over{k_x + \delta k_x (\epsilon) + k_{x0}}}\right ]$ and $\theta_2 = arctan \left [ {{k_y + \delta k_y (\epsilon)  - k_{y0}}\over{k_x + \delta k_x (\epsilon) - k_{x0}}}\right ]$. This change in the eigenspinors rotates the spin polarization in the channel when it is strained. That effect is exploited to implement a SpinFET.

\begin{figure}[!t]
\centering
\includegraphics[width = 0.46\textwidth]{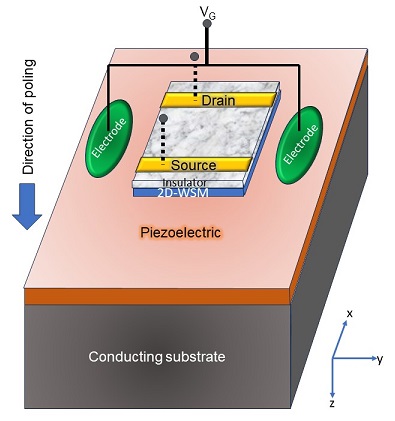}
\caption{Structure of the SpinFET based on a strained two-dimensional Weyl semimetal.}
\label{fig:transistor}
\end{figure}

The SpinFET structure is shown in Fig. \ref{fig:transistor}. It has a two-dimensional layer of a Weyl semimetal (2D-WSM) on a piezoelectric thin film that has been deposited on a conducting substrate. A thin insulating capping layer is placed on the 2D-WSM. This structure is etched into a mesa. Two electrodes are placed on the piezoelectric layer flanking the mesa and they are electrically shorted together. Two ferromagnetic contacts are delineated on top of the insulating thin film and diffused down to contact the 2D-WSM layer underneath. These two ferromagnetic contacts act as the source and the drain of the transistor structure discussed here and the two electrically shorted contacts together act as the gate contact. The 2D-WSM layer acts as the channel material. 

The piezoelectric is poled in the vertical direction. If a (gate) voltage is applied between the shorted electrode pair and the back of the (grounded) conducting substrate, then most of it is dropped across the piezoelectric since the substrate is conducting. This generates a strain in the region of the piezoelectric pinched between the two electrodes, which is transferred to the 2D-WSM. The strain is mostly biaxial \cite{ten} and the component of strain along the line joining the two electrodes will be tensile if the gate voltage polarity is such that the resulting electric field is along the direction of poling. In that case, the strain component in the transverse direction, i.e. the direction of current flow between the source and the drain, will be compressive. Such strain will change the expressions for the thetas and hence the eigenspinors as stated earlier, which will make the spin polarization of carriers in the channel rotate. Since the source and the drain act as spin polarizer and analyzer, this rotation will change the channel conductance just like in an ordinary SpinFET \cite{datta,bandy}. That implements the transistor. The only difference with the ordinary SpinFET is that here the gate voltage does not modulate the spin-orbit interaction in the channel to rotate the spin, but instead modulates the strain in the channel to rotate the spin.

We will assume first that the 2D-WSM in Fig. \ref{fig:transistor} is semi-infinite in the $y$-direction. The Hamiltonian is then invariant in $y$ and therefore both $k_y$ and $k_y + \delta k_y(\epsilon) $ are good quantum numbers. From Equation (\ref{new-dispersion}), it is clear that for any given energy $E$ and given transverse wave vector component $k_y^{\prime}  = k_y + \delta k_y(\epsilon) $, the $k_x^{\prime} = k_x + \delta k_x(\epsilon) $ values for the positively traveling states associated with the two different chiralities (or Weyl nodes) will be different and related according to the relation 
\begin{eqnarray}
    E & = & \hbar v \sqrt{\left (k_{x1}^{\prime} + k_{x0} \right )^2 + \left (k_y^{\prime} + k_{y0} \right )^2} \nonumber \\
    & = & \hbar v \sqrt{\left (k_{x2}^{\prime} - k_{x0} \right )^2 + \left (k_y^{\prime} - k_{y0} \right )^2}, 
\end{eqnarray}
where the subscripts `1' and `2' are associated with the two different chiralities or Weyl nodes. This yields
\begin{eqnarray}
     k_{x1}^{\prime}  \left (E, k_y + \delta k_y (\epsilon) \right ) 
    & = & \sqrt{(E/\hbar v)^2 - \left (k_y^{\prime}(\epsilon) + k_{y0} \right )^2} - k_{x0} \nonumber \\
     k_{x2}^{\prime}  \left (E, k_y + \delta k_y(\epsilon) \right )
     & = & \sqrt{(E/\hbar v)^2 - \left (k_y^{\prime}(\epsilon) - k_{y0} \right )^2} + k_{x0} \nonumber \\
    \label{wavevectors}
\end{eqnarray}
The expressions for the theta’s will then be modified to 
\begin{eqnarray}
    \theta_1^{\prime} \left (E, k_y + \delta k_y(\epsilon) \right )  =   arctan \left [ {{k_y^{\prime} + k_{y0}}\over{k_{x1}^{\prime} \left (E, k_y^{\prime} \right )+ k_{x0}}}\right ] \nonumber  \\
   =  arctan \left [ {{k_y + \delta k_y(\epsilon)  + k_{y0}}\over{\sqrt{(E/\hbar v)^2 - \left (k_y + \delta k_y(\epsilon) + k_{y0} \right )^2}}}\right ] 
    \end{eqnarray}
    and 
\begin{eqnarray}
    \theta_2^{\prime} \left (E, k_y + \delta k_y(\epsilon) \right )  =   arctan \left [ {{k_y^{\prime} (\epsilon) - k_{y0}}\over{k_{x2}^{\prime} \left (E, k_y^{\prime}(\epsilon) \right )- k_{x0}}}\right ] \nonumber \\
    =   arctan \left [ {{k_y + \delta k_y(\epsilon) - k_{y0}}\over{\sqrt{(E/\hbar v)^2 - \left (k_y + \delta k_y(\epsilon) - k_{y0} \right )^2}}}\right ].
    \end{eqnarray}

Let us assume that the ferromagnetic source and the drain contacts are both magnetized in the +$y$-direction as shown in Fig. \ref{fig:transistor}. We will assume that the source injects only +$y$-polarized spins into the 2D-WSM at the complete exclusion of –$y$-polarized spins (perfect spin polarizer). This assumption can be relaxed \cite{eleven}, but it is not necessary at this point to explain how the SpinFET works. An injected +$y$-polarized spin will couple into the two eigenspin states in the 2D-WSM with (wave vector dependent) coupling coefficients $C_1$ and $C_2$. We can view this occurrence as the incident +$y$-polarized beam splitting into two forward traveling beams, each corresponding to an eigenspinor in the 2D-WSM channel. These two beams propagate in different directions for any given energy $E$ and transverse wave vector $k_y$. Hence the 2D-WSM channel behaves like a birefringent medium \cite{eleven}. The beam splitting is expressed by the equation 
\begin{eqnarray}
{{1}\over{\sqrt{2}}} \left [
\begin{array}{c}
1 \\
i
\end{array}
\right ]
& = & C_1 \psi_{\uparrow}^{(1)} + C_2 \psi_{\uparrow}^{(2)} 
=  C_1\left [
 \begin{array}{c}
 1 \\
 e^{i \theta_1^{\prime}}
 \end{array}
 \right ] 
 +
 C_2 \left [
 \begin{array}{c}
 1 \\
 e^{i \theta_2^{\prime}}
 \end{array}
 \right ] \nonumber \\
({\bf +y-pol}) &&
 \label{spinors}
 \end{eqnarray}

The coupling coefficients can be found from Equation (\ref{spinors}) as 
\begin{eqnarray}
    C_1 & = & {{1}\over{\sqrt{2}}} {{e^{i \theta_2^{\prime}} - e^{i \pi /2}}\over{e^{i \theta_2^{\prime}} - e^{i \theta_1^{\prime}}}} \nonumber \\
    C_2 & = & {{1}\over{\sqrt{2}}} {{e^{i \theta_1^{\prime}} - e^{i \pi /2}}\over{e^{i \theta_1^{\prime}} - e^{i \theta_2^{\prime}}}}
\end{eqnarray}

In the drain contact, the two beams interfere. The phase difference between them (accrued in traversing the channel) determine the spinor wavefunction (and hence the spin polarization) of the electron impinging on the drain. This, in turn, determines the transmission probability through the drain contact (spin analyzer) and therefore the source-to-drain current. We will show that the phase difference can be altered with the gate potential which strains the 2D-WSM. Ultimately, this elicits the SpinFET functionality.

The spinor wavefunction at the drain end is given by 
\begin{eqnarray}
    [ \psi]_{drain} & = &  C_1
    \left [ 
    \begin{array}{c}
 1 \\
 e^{i \theta_1^{\prime}}
 \end{array} \right ]
 e^{i \left (k_{x1}^{\prime}L + k_y^{\prime}W \right )}
 + C_2
 \left [ \begin{array}{c}
 1 \\
 e^{i \theta_2^{\prime}}
 \end{array}
 \right ]
 e^{i \left (k_{x2}^{\prime}L + k_y^{\prime}W \right )} \nonumber \\
 & = & 
 {{1}\over{\sqrt{2}}} {{e^{i \theta_2^{\prime}} - e^{i \pi /2}}\over{e^{i \theta_2^{\prime}} - e^{i \theta_1^{\prime}}}}
 \left [ 
    \begin{array}{c}
 1 \\
 e^{i \theta_1^{\prime}}
 \end{array} \right ]
 e^{i \left (k_{x1}^{\prime}L + k_y^{\prime}W \right )}
  \nonumber \\
  & + & {{1}\over{\sqrt{2}}} {{e^{i \theta_1^{\prime}} - e^{i \pi /2}}\over{e^{i \theta_1^{\prime}} - e^{i \theta_2^{\prime}}}}
 \left [ 
    \begin{array}{c}
 1 \\
 e^{i \theta_2^{\prime}}
 \end{array} \right ]
 e^{i \left (k_{x2}^{\prime}L + k_y^{\prime}W \right )},
\end{eqnarray}
where $L$ is the channel length (distance between source and drain contacts) and $W$ is the transverse displacement of the electron (in the $y$-direction) as it traverses the channel. Neglecting multiple reflection effects, the amplitude of transmission through the drain contact, $t$, is the projection of the arriving spinor $[ \psi]_{drain}$ on the drain’s spin polarization (which is the +$y$-polarization). Hence
\begin{eqnarray}
    t & = & {{1}\over{2}}e^{i k_y^{\prime} W}
    \left [ 
    \begin{array}{c c}
    1 & i 
    \end{array}
    \right ] \nonumber \\
    & \times &
    \left (
   {{e^{i \theta_2^{\prime}} - e^{i \pi /2}}\over{e^{i \theta_2^{\prime}} - e^{i \theta_1^{\prime}}}}
 \left [ 
    \begin{array}{c}
 1 \\
 e^{i \theta_1^{\prime}}
 \end{array} \right ]
 e^{i k_{x1}^{\prime}L}
 +
 {{e^{i \theta_1^{\prime}} - e^{i \pi /2}}\over{e^{i \theta_1^{\prime}} - e^{i \theta_2^{\prime}}}}
 \left [ 
    \begin{array}{c}
 1 \\
 e^{i \theta_2^{\prime}}
 \end{array} \right ]
 e^{i k_{x2}^{\prime}L}
 \right ) \nonumber \\
 & = & 
 {{1}\over{2}}e^{i \left ( k_y^{\prime} W + k_{x1}^{\prime}L \right )}{{e^{i \theta_1^{\prime}} + e^{i \theta_2^{\prime}} + e^{i \left ( \theta_1^{\prime} + \theta_2^{\prime} + \pi/2 \right )} -e ^{i \pi/2}}\over{e^{i \theta_2^{\prime}} - e^{i \theta_1^{\prime}}}} \left (1 - e^{i \phi} \right ) \nonumber \\
 \label{amplitude}
\end{eqnarray}
where $\phi = \left ( k_{x2}^{\prime} - k_{x1}^{\prime} \right )L$.

Using Equation (\ref{wavevectors}), the phase difference $\phi$ can be written as
\begin{eqnarray}
\phi \left (E, k_y^{\prime} \right ) & = & \phi \left (E, k_y + \delta k_y(\epsilon) \right ) \nonumber \\
& = &  (\sqrt{\left ( {{E}\over{\hbar v}} \right )^2 - \left (k_y + \delta k_y(\epsilon) - k_{y0} \right )^2}  \nonumber \\
& - &  \sqrt{\left ( {{E}\over{\hbar v}} \right )^2- \left (k_y + \delta k_y(\epsilon) + k_{y0} \right )^2} + 2k_{x0} )L.
\nonumber \\
\label{phi-relation}
\end{eqnarray}

Curiously, the phase difference is independent of both $E$ and $k_y^{\prime}$ [and is approximately equal to $2k_{x0}L$] for large transverse wave vector component $k_y^{\prime} \gg k_{y0}$ and also small transverse wave vector component $k_y^{\prime} \ll k_{y0}$ . This has a detrimental effect on transistor operation since a gate-voltage-independent phase shift does not allow gate control of the source-to-drain current that underlies the transistor action. Hence we expect the conductance modulation (or the conductance on/off ratio) to be poor. This is typical of virtually all two-dimensional SpinFETs \cite{one,eleven}, which is why they are not suitable for use as digital switches.

Finally, the probability of transmission through the drain contact is obtained from Equation (\ref{amplitude}) as
\begin{eqnarray}
    T \left (E, k_y^{\prime}\right ) & = & |t\left (E, k_y^{\prime} \right ) |^2 = A \left (E, k_y^{\prime} \right ) sin^2 \left [ \phi\left (E, k_y^{\prime} \right )/2  \right ] \nonumber \\
    & = & A \left (E, k_y + \delta k_y (\epsilon) \right ) sin^2 \left [ \phi\left (E, k_y + \delta k_y (\epsilon) \right )/2  \right ], \nonumber \\
\label{transmission}
\end{eqnarray}
where $A \left (E, k_y + \delta k_y(\epsilon) \right ) = \left | {{e^{i \theta_1^{\prime}} + e^{i \theta_2^{\prime}} + e^{i \left ( \theta_1^{\prime} + \theta_2^{\prime} + \pi/2 \right )} -e ^{i \pi/2}}\over{e^{i \theta_2^{\prime}} - e^{i \theta_1^{\prime}}}} \right |^2$. 

It is obvious from Equation (\ref{transmission}) that the strain $\epsilon$ generated by the gate voltage $V_G$ can change the transmission probability $T \left (E, k_y^{\prime}\right ) = T\left (E, k_y + \delta k_y(\epsilon) \right ) = T\left (E, k_y + \delta k_y \left ( V_G \right ) \right ) $ through the 2D-WSM channel and hence the source-to-drain channel conductance $G_{SD}$ in the linear response regime given by the Landauer-B\"uttiker formula \cite{Springer}
\begin{equation}
G_{SD} \left (V_G \right )  =  {{q^2 W_y}\over{\pi h}}  \int_0^{\infty} dE \int dk_y T \left ( E, k_y + \delta k_y \left ( V_G \right )\right ) \left [ - {{\partial f(E)}\over{\partial E}}\right ] ,
\label{Landauer}
\end{equation}
where $q$ is the electronic charge, $h$ is the Planck constant and $f(E)$ is the Fermi-Dirac occupation probability. Using Equation (\ref{transmission}) in (\ref{Landauer}), we obtain 
\begin{eqnarray}
    G_{SD} \left (V_G \right ) & = &  {{q^2 W_y}\over{\pi h}} \nonumber \\
     & \times &  \int_0^{\infty} dE \int dk_y  \{ A \left (E, k_y + \delta k_y \left (V_G \right ) \right )  \nonumber \\
     & \times & sin^2 \left [ \phi\left (E, k_y + \delta k_y \left ( V_G \right )  \right )/2  \right ] 
      \left [ - {{\partial f(E)}\over{\partial E}}\right ]  \}. \nonumber \\
\end{eqnarray}
At low temperatures, the above equation simplifies to 
\begin{eqnarray}
G_{SD} \left (V_G \right ) & = & {{q^2 W_y}\over{\pi h}} \nonumber \\
& \times & \int_0^{k_F} dk_y \{  A \left (E_F, k_y + \delta k_y \left( V_G \right)  \right )  \nonumber \\
& \times &  sin^2 \left [ \phi\left (E_F, k_y + \delta k_y \left( V_G \right)  \right ) /2 \right ] \}, 
\label{linear-response}
\end{eqnarray}
where $E_F$ is the Fermi energy, $k_F$ is the Fermi wave vector and $V_G$ is the gate voltage. 

In order to illustrate SpinFET action, we will make the following assumption. We will assume that $\delta k_y$ increases linearly with strain. Hence $\delta k_y = \lambda \epsilon$, where $\lambda$ is a constant and $\epsilon$ is the strain. From the results in Ref. [\onlinecite{three}], we estimate $\lambda$ $\sim$ 500 m$^{-1}$/microstrain.

\begin{figure}[!t]
\centering
\includegraphics[width = 0.46\textwidth]{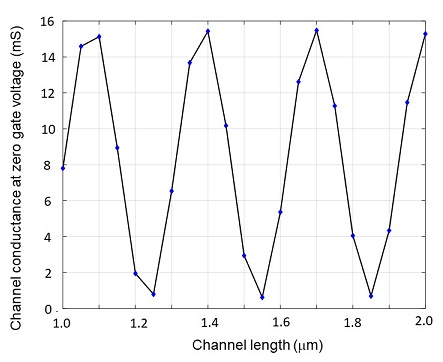}
\caption{Source-to-drain conductance of the SpinFET as a function of channel length. The material and structural parameters used for this simulation are given in the text. }
\label{fig:channellength}
\end{figure}

The strain generated in a piezoelectric by an applied electric field ${\cal E}$ is $\epsilon = d_{33} {\cal E}$, where $d_{33}$ is the diagonal element of the piezoelectric tensor. The electric field is related to the gate voltage according to ${\cal E} = V_G/d$ where $d$ is the thickness of the piezoelectric layer. We will assume that the piezoelectric is a two-dimensional piezoelectric like CrTe$_2$ with a layer thickness of 1 nm and $d_{33}$ = 17 pC/N \cite{2Dpiezo1, 2Dpiezo2}.
We will further assume that the strain is fully transferred from the piezoelectric to the 2D-WSM. Hence, the gate-voltage-generated strain in the 2D-WSM is $\epsilon = d_{33} V_G/d$. Therefore, we get that $\delta k_y = \lambda  d_{33} V_G/d$.

We can use Equation (\ref{linear-response}) to obtain the low temperature linear response source to drain conductance of the transistor in Fig. \ref{fig:transistor} as a function of the gate voltage $V_G$. For numerical results, we make the following arbitrary but reasonable assumptions: $E_F$ = 10 meV,   $v$ = 10$^5$ m/sec,  $k_{x0} = k_{y0}$ = 10$^7$ m$^{-1}$, $k_F \sim E_F/\left ( \hbar v_0 \right ) - k_{y0}$ = 6$\times$10$^6$ m$^{-1}$, and $W_y$ = 1 $\mu$m. 

In Fig. \ref{fig:channellength}, we plot the channel conductance $G_{SD}$ at zero gate voltage as a function of the channel length $L$. Interestingly, the channel conductance {\it does not decrease monotonically with increasing channel length}, but instead shows oscillations, which is a non-classical behavior. This is a consequence of quantum interference between the two eigenspin states that determines the channel conductance. The phase difference between the two states $\phi$ increases with $L$ and the channel conductance has an oscillatory dependence on $\phi$ as can be seen from Equation (\ref{linear-response}). This is manifested in Fig. \ref{fig:channellength}.

\begin{figure}[!h]
\centering
\includegraphics[width = 0.46\textwidth]{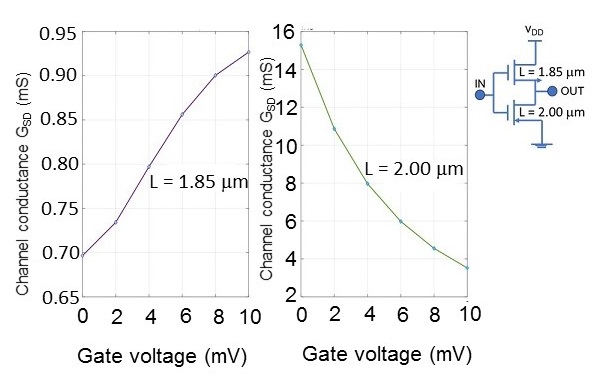}
\caption{Transfer characteristics of the SpinFET in the range 0 - 10 mV of gate voltage for two different channel lengths of 1.85 and 2 $\mu$m. In one case, the transconductance (slope) is positive, and in the other case, it is negative. Two such devices can be connected in series to elicit complementary operation, as in CMOS. This is shown in the inset.}
\label{fig:transfer}
\end{figure}

Finally, we plot the transfer characteristics of the SpinFET, i.e. channel conductance versus gate voltage, for two different values of the channel length, in Fig. \ref{fig:transfer}. The two channel lengths are picked from Fig. \ref{fig:channellength} such that one corresponds to a peak and the other to a trough of the conductance plot. Note that in one case, the channel conductance increases with gate voltage resulting in a positive slope or positive transconductance. In the other case, the slope is negative meaning that the transconductance has a negative sign. This change in sign can be exploited to elicit complementary operation as in a complementary metal oxide semiconductor field effect transistor (CMOS) made of a series combination of a p-channel transistor with negative transconductance and an n-channel transistor with positive transconductance. Here, we can place two SpinFETs of slightly different channel lengths in series as shown in the inset of the figure to ensure that appreciable current flows only during switching. This reduces standby power dissipation. Note also that this feature can be manifested over a small gate voltage range of 0 - 10 mV, implying that this is intrinsically a low power device.

Before concluding, we point out that in this work, we have made many approximations (linear response, low temperature, etc.) in order to elucidate the physics behind this device. Relaxation of these approximations will surely result in quantitative changes, but possibly not qualitative changes. We also mention that since semimetals are fairly conductive, space charge effects are weak or non-existent, which obviates the need for self consistent solutions (involving the Poisson equation) to find the transmission probability and hence the linear response source-to-drain conductance as a function of the gate voltage, i.e. the transfer characteristic of the transistor. 

Curiosities such as these are of course not intended to compete with the silicon transistor juggernaut for mainstream applications. Even the ideal original SpinFET was shown to have inferior characteristics compared to ordinary silicon transistors for digital applications \cite{cahay}. Nonetheless, the unusual properties stemming from quantum mechanics, such as the transconductance changing sign when the channel length is varied slightly, may find niche applications. Just as the original SpinFET idea stimulated vigorous research in applied spintronics, we hope that these explorations will inspire research in electronic device applications of emerging quantum materials like topological insulators and Weyl semimetals, whose device applications are rarely explored. 

\smallskip

\noindent{\bf Acknowledgement}: This work was supported by the Virginia Commonwealth University QUEST fund and a Virginia Microelectronics Corporation grant.

\end{document}